\documentclass[prc,aps,superscriptaddress,twocolumn,showpacs,nofootinbib]{revtex4}

\usepackage{graphicx,amsmath,amssymb,bm,multirow}
\usepackage{amsthm,mathrsfs}
\usepackage{amsfonts}
\usepackage{dcolumn}
\usepackage{bm}
\usepackage[usenames,dvipsnames]{color}
\graphicspath{{figures/}}

\newcommand{\br}{{\vec{r}}}

\newcommand{\beq}{\begin{equation}}
\newcommand{\eeq}{\end{equation}}
\newcommand{\beqn}{\begin{eqnarray}}
\newcommand{\eeqn}{\end{eqnarray}}

\newcommand{\bsub}{ \begin{subequations}}
\newcommand{\esub}{ \end{subequations}}

\renewcommand{\vec}[1]{\mbox{\boldmath $#1$}}

\begin{document}
\title{Low-energy hypernuclear spectra with microscopic particle-rotor model
with relativistic point coupling hyperon-nucleon interaction}
\author{H. Mei}
\affiliation{Department of Physics, Tohoku University, Sendai 980-8578,Japan}
\affiliation{School of Physical Science and Technology,
             Southwest University, Chongqing 400715, China}
\author{K. Hagino}
\affiliation{Department of Physics, Tohoku University, Sendai 980-8578,Japan}
\affiliation{Research Center for Electron Photon Science, Tohoku University, 1-2-1 Mikamine, Sendai 982-0826, Japan}
\affiliation{
National Astronomical Observatory of Japan, 2-21-1 Osawa,
Mitaka, Tokyo 181-8588, Japan}

\author{J. M. Yao}
\affiliation{Department of Physics and Astronomy, North Carolina University, Chape Hill 27599-3255, USA}
\affiliation{School of Physical Science and Technology,
             Southwest University, Chongqing 400715, China}
\author{T. Motoba}
\affiliation{Laboratory of Physics, Osaka Electro-Communications University,
             Neyagawa 572-8530, Japan}
\affiliation{Yukawa Institute for Theoretical Physics, Kyoto University, Kyoto 606-8502, Japan }

\begin{abstract}
We extend the microscopic particle-rotor model  for hypernuclear
low-lying states by including the derivative  and  tensor coupling
terms in
the point-coupling nucleon-$\Lambda$ particle ($N\Lambda$) interaction.
Taking $^{13}_{~\Lambda}$C as an example,
we show that
a good overall description for excitation spectra is achieved with  four sets of effective
$N\Lambda$ interaction.
We find  that the $\Lambda$ hyperon binding energy decreases monotonically with increasing
the strengths of the high-order interaction terms. In particular, the tensor coupling term
decreases the energy splitting between the first $1/2^-$ and $3/2^-$ states and increases
the energy splitting between the first $3/2^+$ and $5/2^+$ states in $^{13}_{~\Lambda}$C.
\end{abstract}

\maketitle

\section{Introduction}

The spectroscopic data on low-lying states of light $\Lambda$ hypernuclei have been
accumulated~\cite{Hashimoto06}
and more data on those of medium and heavy hypernuclei are expected to be obtained with the
next-generation facilities such as J-PARC~\cite{Tamura09}.
Rich information on the hyperon-nucleon interaction in nuclear medium and the impurity effect of a $\Lambda$
particle on nuclear structure are contained in these data.
Because hyperon-nucleon and hyperon-hyperon scattering experiments are difficult to perform,
the structure of hypernuclei has been playing a
vital role in order to shed light on baryon-baryon interactions.
Such information is crucial in order to understand also neutron stars, in which
hyperons may emerge in the inner part~\cite{Glendenning00}.
However, extracting information on baryon-baryon interactions from the spectroscopic data relies
much on nuclear models.

In the past decades, several different types of theoretical models have been developed to
study the structure of hypernuclei,
including an ab-initio method~\cite{abinitio}, a cluster
model~\cite{Motoba83,Hiyama99,Bando90,Hiyama03,Cravo02, Suslov04, Shoeb09},
a shell model~\cite{Dalitz78,Gal71,Millener}, the anti-symmetrized molecular
dynamics (AMD)~\cite{Isaka11,Isaka11-2,Isaka12,Isaka13},
self-consistent mean-field approach~\cite{Zhou07,Win08,Schulze10,Win11,Lu11,Weixia14,Li13,Lu14,HY14}
and the generator coordinator method (GCM) based on energy density functionals~\cite{Mei15-2}.
In recent years, we have also developed a microscopic particle rotor model (MPRM) for
hypernuclear low-lying states based on the beyond-mean-field approach~\cite{Mei2014,Mei2015}.
In contrast to the GCM for the whole hypernuclei~\cite{Mei15-2}, where the wave function of the
hypernuclear states is given as a superposition of hypernuclear mean-field states,  the
hypernuclear states in the MPRM  are constructed by coupling a hyperon to low-lying states of
the core nucleus. The MPRM provides a convenient way to analyze the components of hypernuclear
wave function and has been applied to study the low-lying states
of $^{9}_{\Lambda}$Be~\cite{Mei2014}, $^{13}_{~\Lambda}$C, $^{21}_{~\Lambda}$Ne and $^{155}_{\Lambda}$Sm
hypernuclei~\cite{Mei2015}.
For the sake of simplicity, only the leading-order four-fermion coupling terms of scalar and
vector types were adopted for the
 $N\Lambda$ effective interaction in these studies.

The aim of this paper is to extend the previous calculations by implementing the higher-order
derivative  and  tensor $N\Lambda$ interaction terms
in the point-coupling interaction~\cite{Tanimura2012}.
The derivative terms simulate to some extent the finite-range character of $N\Lambda$
interaction and these terms are expected to be more pronounced in light
hypernuclei~\cite{Hiyama14}. On the other hand, the
tensor $N\Lambda$ interaction is important to reproduce a small hyperon spin-orbit splitting
in $\Lambda$ hypernuclei~\cite{Noble80}.
It is therefore important to assess the effect of these terms on hypernuclear low-lying states.

The paper is organized as follows.
In Sec.~\ref{Sec:framework}, we present the main formulas of the microscopic PRM for $\Lambda$
hypernuclei with the full point-coupling effective $N\Lambda$ interaction.
In Sec.~\ref{Sec:Results}, we show the results for hypernuclear low-lying states in
$^{13}_{~\Lambda}$C and discuss
the influence of the higher-order terms on the energy spectra. We then summarize the paper
in Sec.~\ref{Sec:Summary}.

\section{Microscopic particle-rotor model for $\Lambda$ hypernuclei}
\label{Sec:framework}
In this paper, we consider a single-$\Lambda$ hypernucleus and describe the hypernuclear
low-lying states
using the microscopic particle-rotor model (MPRM).
Since the detailed formulas for the MPRM  have been given in Refs.~\cite{Mei2014,Mei2015},
we give here only the main formulas of this approach.
To this end, we put a particular emphasis  on the implementation of the higher-order $N\Lambda$
interaction terms.

The basic idea of the MPRM is to construct a hypernuclear wave function by coupling the
valence $\Lambda$ hyperon to the low-lying states of nuclear core in the laboratory frame, that is,
\begin{eqnarray}
 \label{wavefunction}
 \displaystyle \Psi_{JM}(\vec{r},\{\vec{r}_i\})
 &=&\sum_{n,j,\ell, I}  {\mathscr R}_{j\ell n I}(r) {\mathscr F}^{JM}_{j\ell n I }(\hat{\vec{r}}, \{\vec{r}_i\}),
\end{eqnarray}
with
\begin{equation}
 {\mathscr F}^{JM}_{j\ell  n I}(\hat{\vec{r}}, \{\vec{r}_i\})
= [{\mathscr Y}_{j\ell }(\hat{\vec{r}})\otimes
\Phi_{n I}(\{\vec{r}_i\})]^{(JM)},
\end{equation}
where $\vec{r}$ and $\vec{r}_i$ are the coordinates of the $\Lambda$ hyperon and the
nucleons, respectively. Here, $J$ is the angular momentum
for the whole system, while $M$ is its
projection onto the $z$-axis in the laboratory frame.
${\mathscr Y}_{j\ell}(\hat{\vec{r}})$ is the spin-angular wave function for the $\Lambda$ hyperon.
$\vert\Phi_{n I}\rangle$ is the wave functions for the low-lying states
of the core nucleus, where $I$ represents the angular momentum
of the core state and $n=1, 2, \ldots$ distinguish
different core states with the same angular momentum $I$.
In the MPRM, the core states $\vert\Phi_{nI}\rangle$  are constructed with the quantum-number
projected GCM approach~\cite{Mei2014,Mei2015}.
For convenience, hereafter we introduce the
shorthanded notation $k=\{j\ell n I\}$ to represent different channels.

In Eq. (\ref{wavefunction}), ${\mathscr R}_{k}(r)$ is the radial wave function for the
$\Lambda$-particle.
In the relativistic approach, it is given as a four-component Dirac spinor
\begin{equation}
\label{wavefunction2}
{\mathscr R}_{k}(r)=
\begin{pmatrix}
f_{k}(r)\\
i g_{k}(r)\vec{\sigma} \cdot \hat{\vec{r}}
\end{pmatrix}.
\end{equation}

We assume that the Hamiltonian $\hat H$ for the whole $\Lambda$ hypernucleus is given as
\begin{eqnarray}
\hat H =\hat T_\Lambda +  \hat H_{\rm c}+ \sum^{A_c}_{i=1} \hat{V}^{N\Lambda}(\vec{r},\vec{r}_{i}).
\label{eq:H}
\end{eqnarray}
Here $\displaystyle\hat T_\Lambda=-i\vec{\alpha}\cdot\nabla_{\Lambda}+\gamma^0 m_{\Lambda}$ is the
relativistic kinetic energy of $\Lambda$ hyperon,
where $m_{\Lambda}$ is the mass of $\Lambda$ particle, and $\vec{\alpha}$ and $\gamma^0$ are
the Dirac matrices.
$\hat H_c$ is the many-body Hamiltonian for the core nucleus~\cite{Buvenich02}, with which the
core state $\vert\Phi_{n I}\rangle$  satisfies
$\hat H_c \vert\Phi_{n I}\rangle= E_{n I} \vert\Phi_{n I}\rangle$.
The last term in Eq. (\ref{eq:H}) represents the $N\Lambda$ interaction term between the
valence $\Lambda$ particle and the nucleons in the core nucleus, where $A_c$ is the mass number
of the core nucleus.

We construct the $N\Lambda$ interaction $\hat{V}^{N\Lambda}$ based on  the relativistic
point-coupling model~\cite{Tanimura2012},
in which the
energy functional for the $N\Lambda$ interaction reads
\begin{align}
\label{EDF}
E^{(N\Lambda)}_{\rm int}=&\int d\vec{r}
\Big[ \alpha_S^{N\Lambda}\rho_S(\vec{r}) \rho^\Lambda_S(\vec{r})
+\alpha_V^{N\Lambda}\rho_V(\vec{r}) \rho^\Lambda_V(\vec{r})
\nonumber \\ &
+\delta_S^{N\Lambda}\rho_S(\vec{r}) \Delta\rho^\Lambda_S(\vec{r})
+\delta_V^{N\Lambda}\rho_V(\vec{r}) \Delta\rho^\Lambda_V(\vec{r})
\nonumber \\&
+\alpha_T^{N\Lambda} \rho^\Lambda_T(\vec{r}) \rho_V(\vec{r}) \Big].
\end{align}
Here $\rho_S$, $\rho_V$ and $\rho_T$ are the scalar, the vector and the tensor densities
defined in Ref.~\cite{Tanimura2012}, respectively. Taking the second functional derivative of
Eq. (\ref{EDF}) with respect to the densities~\cite{Ring80},
\begin{align}
\hat{V}^{N\Lambda}(\vec{r},\vec{r}_i)=&
\frac{\delta^2 E[\rho]}{\delta \rho^\Lambda_S(\vec{r}) \delta\rho_S(\vec{r}_i)}
+\frac{\delta^2 E[\rho]}{\delta \rho^\Lambda_V(\vec{r}) \delta\rho_V(\vec{r}_i)} \nonumber\\
&+\frac{\delta^2 E[\rho]}{\delta \rho^\Lambda_T(\vec{r}) \delta\rho_V(\vec{r}_i)},
\end{align}
we obtain the following form for the $N\Lambda$ effective interaction
\beq
\hat{V}^{N\Lambda} = \hat{V}^{N\Lambda}_{\rm S} + \hat{V}^{N\Lambda}_{\rm V}
+ \hat{V}^{N\Lambda}_{\rm Ten},
\eeq
where the scalar, vector and tensor types of coupling terms read
\begin{align}
\label{Scalar}
\hat{V}^{N\Lambda}_{\rm S}(\vec{r},\vec{r}_i)=& \alpha_S^{N\Lambda} \gamma^0_\Lambda
\delta(\vec{r}-\vec{r}_i)\gamma^0_N
+\delta_S^{N\Lambda}\gamma^0_\Lambda
\Big[\overleftarrow{\nabla}^2 \delta(\vec{r}-\vec{r}_i) \nonumber \\
~&+ \delta(\vec{r}-\vec{r}_i)\overrightarrow{\nabla}^2+ 2 \overleftarrow{\nabla}
\cdot\delta(\vec{r}-\vec{r}_i)
 \overrightarrow{\nabla}\Big]\gamma^0_N  \\
 \label{Vector}
\hat{V}^{N\Lambda}_{\rm V}(\vec{r},\vec{r}_i)=&\alpha_V^{N\Lambda} \delta(\vec{r}-\vec{r}_i)
+\delta_V^{N\Lambda} \Big[\overleftarrow{\nabla}^2 \delta(\vec{r}-\vec{r}_i)
 \nonumber \\
~&+ \delta(\vec{r}-\vec{r}_i) \overrightarrow{\nabla}^2+ 2
\overleftarrow{\nabla}\cdot\delta(\vec{r}-\vec{r}_i) \overrightarrow{\nabla}\Big]
 \\
 \label{Tensor}
\hat{V}^{N\Lambda}_{\rm Ten}(\vec{r},\vec{r}_i)=&
i\alpha_T^{N\Lambda}\gamma^0_\Lambda\Big[\overleftarrow{\nabla} \delta(\vec{r}-\vec{r}_i)
+\delta(\vec{r}-\vec{r}_i)\overrightarrow{\nabla}\Big]\cdot \vec{\alpha}.
\end{align}
Here,$\overrightarrow{\nabla}$ and $\overleftarrow{\nabla}$ are understood to act on the
right and left hands sides of the $\Lambda$ hyperon coordinates, respectively.
Vice versa, Eq. (\ref{EDF}) can be obtained from the above effective $N\Lambda$ interaction
(see Appendix A).
We note that similar terms appear in the chiral hyperon-nucleon interaction~\cite{Polinder06},
in which
the non-derivative four-fermion coupling corresponds to the contact leading-order (LO) term.

With Eqs. (\ref{wavefunction}) and (\ref{eq:H}), the radial wave function ${\mathscr R}_{k}(r)$
in Eq. (\ref{wavefunction2})
and the energy $E_J$ for each hypernuclear state are obtained by solving the following
coupled-channels equations,
\bsub\begin{align}
\label{couple1}
&\left(\frac{d}{dr}-\frac{\kappa-1}{r}\right)g_{k}(r)+(E_{n I}-E_J) f_{k}(r) \nonumber \\
&+\sum_{k'}U^{kk'}_{T}(r) g_{k'}(r)
+\sum_{k'}\left[U^{kk'}_V(r)+ U^{kk'}_S(r)\right] f_{k'}(r)
=0, \\
\label{couple2}
&\left(\frac{d}{dr}+\frac{\kappa+1}{r}\right)f_{k}(r)-(E_{n I}-2 m_{\Lambda}-E_J)g_{k}(r) \nonumber \\
&-\sum_{k'}U^{kk'}_{T}(r) f_{k'}(r)
-\sum_{k'}\left[U^{kk'}_V(r) - U^{kk'}_S(r)\right] g_{k'}(r)
= 0,
\end{align}\esub
where $\kappa$ is defined as  $\kappa=(-1)^{j+\ell+1/2}(j+1/2)$.
The  coupling potentials between different channels are given by
\bsub\begin{align}
U^{kk'}_{\mathrm{S}}(r)\equiv&\langle {\mathscr F}^{JM}_{j l n I} |
\sum_{i=1}^{A_c}\hat{V}_{\mathrm{S}}^{N\Lambda}(\vec{r},\vec{r}_i)
|{\mathscr F}^{JM}_{j' l' n' I'}\rangle \\
U^{kk'}_{\mathrm{V}}(r)\equiv&\langle {\mathscr F}^{JM}_{j l n I} |
\sum_{i=1}^{A_c}\hat{V}_{\mathrm{V}}^{N\Lambda}(\vec{r},\vec{r}_i)
|{\mathscr F}^{JM}_{j' l' n' I'}\rangle \\
U^{kk'}_{T}(r)\equiv&\langle {\mathscr F}^{JM}_{jlnI} |\sum_{i=1}^{A_c}
\hat{V}_{\mathrm{T}}^{N\Lambda}(\vec{r},\vec{r}_i)\cdot
\vec{\sigma}|{\mathscr F}^{JM}_{j'\tilde{l}'n'I'}\rangle,
\end{align}\esub
with
\beq
\hat{V}^{N\Lambda}_{\mathrm{T}} \equiv \alpha_T^{N\Lambda}\left[\overleftarrow{\nabla}
\delta(\vec{r}-\vec{r}_i)+\delta(\vec{r}-\vec{r}_i)\overrightarrow{\nabla}\right].
\eeq
By expanding each of the large $f_{k}(r)$ and small $g_{k}(r)$ components of the Dirac spinors,
Eq.(\ref{wavefunction2}),
in terms of the radial function $R_{\alpha l}(r)$ of a spherical harmonic oscillator, that is,
\bsub\begin{align}
f_{k}(r)=&\sum_{\alpha =1}^{f ^{(k)}_{max}}F_{k\alpha } R^{k}_{\alpha  l}(r), \\
g_{k}(r)=&\sum_{\alpha=1}^{g^{(k)}_{max}}G_{k\alpha}R^{k}_{\alpha \tilde{l}}(r),
\end{align}
\esub
with $l=j\pm 1/2$ and $\tilde{l}=j\mp 1/2$,
the coupled-channels equations (\ref{couple1}), (\ref{couple2}) are transformed into a real
symmetric matrix equation,
\beqn
\label{Matrix_eq}
&&\sum_{\alpha',k'}
\begin{pmatrix}
A^{kk'}_{\alpha \alpha '}+V^{kk'}_{\alpha \alpha '}+S^{kk'}_{\alpha \alpha '} &
B^{kk'}_{\alpha\alpha'}+T^{kk'}_{\alpha\alpha'} \\
B^{kk'}_{\alpha\alpha '}+T^{kk'}_{\alpha\alpha'} & C^{kk'}_{\alpha\alpha'}
+V^{kk'}_{\alpha\alpha'}-S^{kk'}_{\alpha\alpha'}
\end{pmatrix}
\begin{pmatrix}
F^{k'}_{\alpha'} \\
G^{k'}_{\alpha'}
\end{pmatrix}\nonumber\\
&&=E_J\begin{pmatrix}
F^{k}_{\alpha } \\
G^{k}_{\alpha}
\end{pmatrix}.
\eeqn
The dimension of the matrix is $\displaystyle\sum_k f ^{(k)}_{max}+g^{(k)}_{max}$, where $k$
represents different channels.
In Eq. (\ref{Matrix_eq}), the matrix elements are given by
\bsub
\label{Matrix}
\begin{align}
A^{kk'}_{\alpha \alpha '} =& \langle R^{k}_{\alpha  l}(r)|E_{n I}| R^{k'}_{\alpha ' l'}(r)\rangle
\delta_{k,k'}  \\
B^{kk'}_{\alpha \alpha'} =& \langle R^{k}_{\alpha  l}(r)|\frac{d}{dr}-\frac{\kappa-1}{r}|
R^{k'}_{\alpha' \tilde{l}'}(r)\rangle \delta_{k,k'}  \\
C^{kk'}_{\alpha\alpha'} =& \langle R^{k}_{\alpha \tilde{l}}(r) |(E_{n I}-2m_{\Lambda})|
R^{k'}_{\alpha' \tilde{l}'}(r)\rangle \delta_{k,k'}
\end{align}
\begin{align}
\label{Vmatrix}
V^{kk'}_{\alpha \alpha '}=&\langle R^{k}_{\alpha  l}(r) |U^{kk'}_{V}(r)|R^{k'}_{\alpha ' l'}(r)
\rangle\nonumber \\
=&(-1)^{j'+I+J} \sum_{\lambda}
\left\{\begin{matrix}
J       &I & j    \\
\lambda &j'  & I'
\end{matrix}\right\}\langle j\ell  || Y_{\lambda } || j'\ell'  \rangle\nonumber \\
~&\times\int r^2 dr \rho^{nI n'I'}_{\lambda,V}(r)\Big\{\alpha_V^{N\Lambda}R^{k}_{\alpha  l}(r)
R^{k'}_{\alpha ' l'}(r) + \delta_V^{N\Lambda}\nonumber \\
~&
\left[\frac{1}{r^2}\frac{d}{dr}(r^2\frac{d}{dr})-\frac{\lambda(\lambda+1)}{r^2}\right]
\left[R^{k}_{\alpha  l}(r)R^{k'}_{\alpha ' l'}(r)\right]
\Big\}
\end{align}
\begin{align}
S^{kk'}_{\alpha \alpha '}=&\langle R^{k}_{\alpha  l}(r) |U^{kk'}_{S}(r)| R^{k'}_{\alpha ' l'}(r)\rangle
\nonumber \\
=&(-1)^{j'+I+J} \sum_{\lambda}
\left\{\begin{matrix}
J       &I & j    \\
\lambda &j'  & I'
\end{matrix}\right\}\langle j\ell  || Y_{\lambda } || j'\ell'  \rangle\nonumber \\
~&\times\int r^2 dr \rho^{nI n'I'}_{\lambda,S}(r)\Big\{\alpha_S^{N\Lambda}R^{k}_{\alpha  l}(r)
R^{k'}_{\alpha ' l'}(r)+ \delta_S^{N\Lambda} \nonumber \\
~&
\left[\frac{1}{r^2}\frac{d}{dr}(r^2\frac{d}{dr})-\frac{\lambda(\lambda+1)}{r^2}\right]
\left[R^{k}_{\alpha  l}(r)R^{k'}_{\alpha ' l'}(r)\right]
\Big\} \label{Smatrix}
\end{align}
\begin{align}
T^{kk'}_{\alpha \alpha'}=& \langle R^{k}_{\alpha  l}(r)|U^{kk'}_{T}(r)| R^{k'}_{\alpha' \tilde{l}'}(r)
\rangle\nonumber \\
=&-\alpha_T^{N\Lambda}(-1)^{j+I'+J} \sum_{\lambda}
\left\{\begin{matrix}
J       &I & j    \\
\lambda &j'  & I'
\end{matrix}\right\}\int r^2 dr \rho^{nI n'I'}_{\lambda,V}(r)
 \nonumber \\
~&\times\Big\{
\Big[\frac{d R^{k}_{\alpha  l}(r)}{dr}+\frac{\kappa+1}{r}R^{k}_{\alpha  l}(r)\Big]
R^{k'}_{\alpha' \tilde{l}'}(r)\langle j\tilde{\ell}  || Y_{\lambda } || j'\tilde{\ell'}  \rangle
 \nonumber \\
~&+\Big[\frac{d R^{k'}_{\alpha' \tilde{l}'}(r)}{dr}-\frac{\kappa'-1}{r}R^{k'}_{\alpha' \tilde{l}'}(r)\Big]
R^{k}_{\alpha l}(r)\langle j\ell  || Y_{\lambda } || j'\ell'  \rangle
\Big\}. \label{Tmatrix}
\end{align}
\esub
See Appendices B and C for the derivation of Eqs. (\ref{Smatrix}) and (\ref{Tmatrix}), respectively.
In Eq. (\ref{Matrix}), $\rho^{nI n'I'}_{\lambda,V}(r)$ and $\rho^{nI n'I'}_{\lambda,S}(r)$
are the reduced vector and scalar transition densities, respectively, between the nuclear
initial state $|\Phi_{n'I'} \rangle$ and the final state $|\Phi_{nI} \rangle$ defined as
\bsub
\label{TD}
\begin{eqnarray}
\label{TD1}
\rho^{nI n'I'}_{\lambda, V}(r) &=&
\langle \Phi_{nI} || \sum\limits_{i=1}^{A_c}
\frac{\delta(r-r_{i})}{r_{i} r}
Y_\lambda(\hat{\vec{r}}_{i})||\Phi_{n'I'} \rangle,
\\
\label{TD2}
\rho^{nI n'I'}_{\lambda, S}(r) &=&
\langle \Phi_{nI} || \sum\limits_{i=1}^{A_c} \gamma^0_i
\frac{\delta(r-r_{i})}{r_{i} r}
Y_\lambda(\hat{\vec{r}}_{i})||\Phi_{n'I'} \rangle.
\end{eqnarray}
\esub
The detailed expressions for the transition densities can be found in Ref.~\cite{Yao15}.

\begin{table*}[hbt]
\tabcolsep=8pt
\caption{The four parameter sets of relativisitc point-coupling $N\Lambda$ interaction
proposed in Ref.~\cite{Tanimura2012}.}
\label{forces}
\begin{center}
 \begin{tabular}{c|c|c|c|c}\hline\hline
                                   & PCY-S1   &  PCY-S2 &  PCY-S3  &  PCY-S4\\ \hline
$\alpha_S^{N\Lambda}$ (MeV$^{-2}$) &$-2.0305\times10^{-4}$  &$-4.2377\times10^{-5}$
&$-2.0197\times10^{-4}$ &$-1.8594\times10^{-4}$\\
$\alpha_V^{N\Lambda}$ (MeV$^{-2}$) &$~~1.6548\times10^{-4}$ &$~~1.4268\times10^{-5}$
&$~~1.6449\times10^{-4}$&$~~1.4981\times10^{-4}$\\
$\delta_S^{N\Lambda}$ (MeV$^{-4}$) &$~~2.2929\times10^{-9}$ &$~~1.2986\times10^{-9}$
&$~~2.3514\times10^{-9}$&$-1.9958\times10^{-10}$\\
$\delta_V^{N\Lambda}$ (MeV$^{-4}$) &$-2.3872\times10^{-9}$  &$-1.3850\times10^{-9}$
&$-2.4993\times10^{-9}$ &$~~~~0$\\
$\alpha_T^{N\Lambda}$ (MeV$^{-3}$) &$-1.0603\times10^{-7}$  &$~~~~0$   &$-4.082\times10^{-9}$
&$-5.5322\times10^{-8}$\\
\hline\hline
 \end{tabular}
\end{center}
\end{table*}

\begin{figure*}[]
  \centering
 \includegraphics[width=14cm]{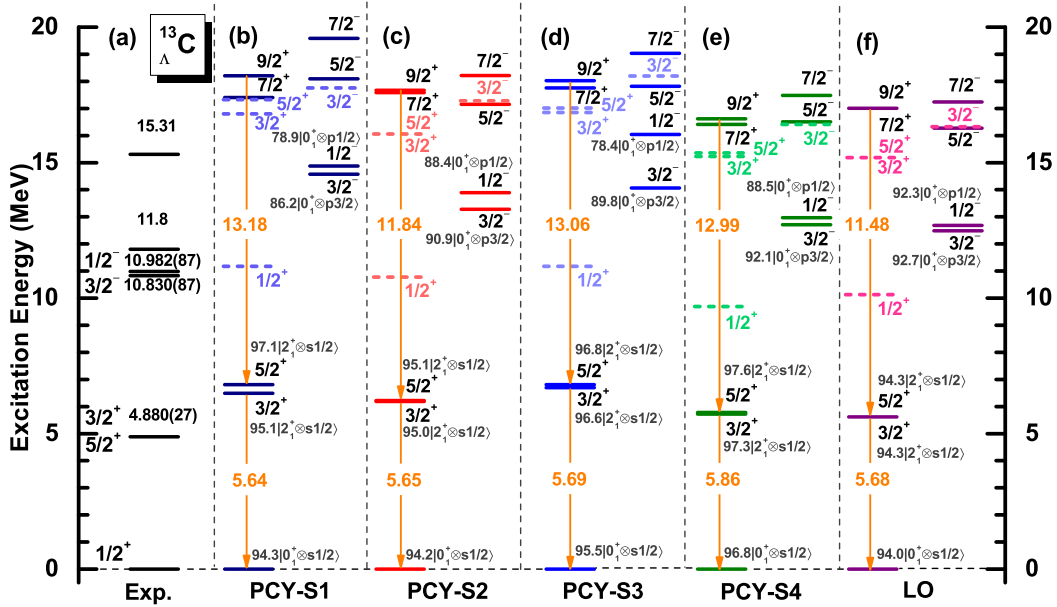}
 \caption{(Color online) The low-energy excitation spectra of $^{13}_{~\Lambda}$C
obtained with the microscopic particle-rotor model with (b)PCY-S1, (c)PCY-S2, (d)PCY-S3 and
(e)PCY-S4.
Fig.~\ref{CSpectrum}(f) shows the spectrum taken from Ref.~\cite{Mei2015}, which was obtained
by  including only the leading-order (LO)
$N\Lambda$ interaction. The experimental data shown in Fig. \ref{CSpectrum}(a) are taken
from Refs. \cite{Ajimura01,Kohri02}.
The numbers with the arrows indicate the B($E2$) value for the $3/2^+_1 \rightarrow 1/2^+_1$
and the $9/2^+_1 \rightarrow 5/2^+_1$ transitions, given in units of $e^2$ fm$^4$.
The dominant component of several hypernuclear states, together with its weight (in percent), is also given.}
  \label{CSpectrum}
\end{figure*}

\begin{figure*}[]
  \centering
 \includegraphics[width=14cm]{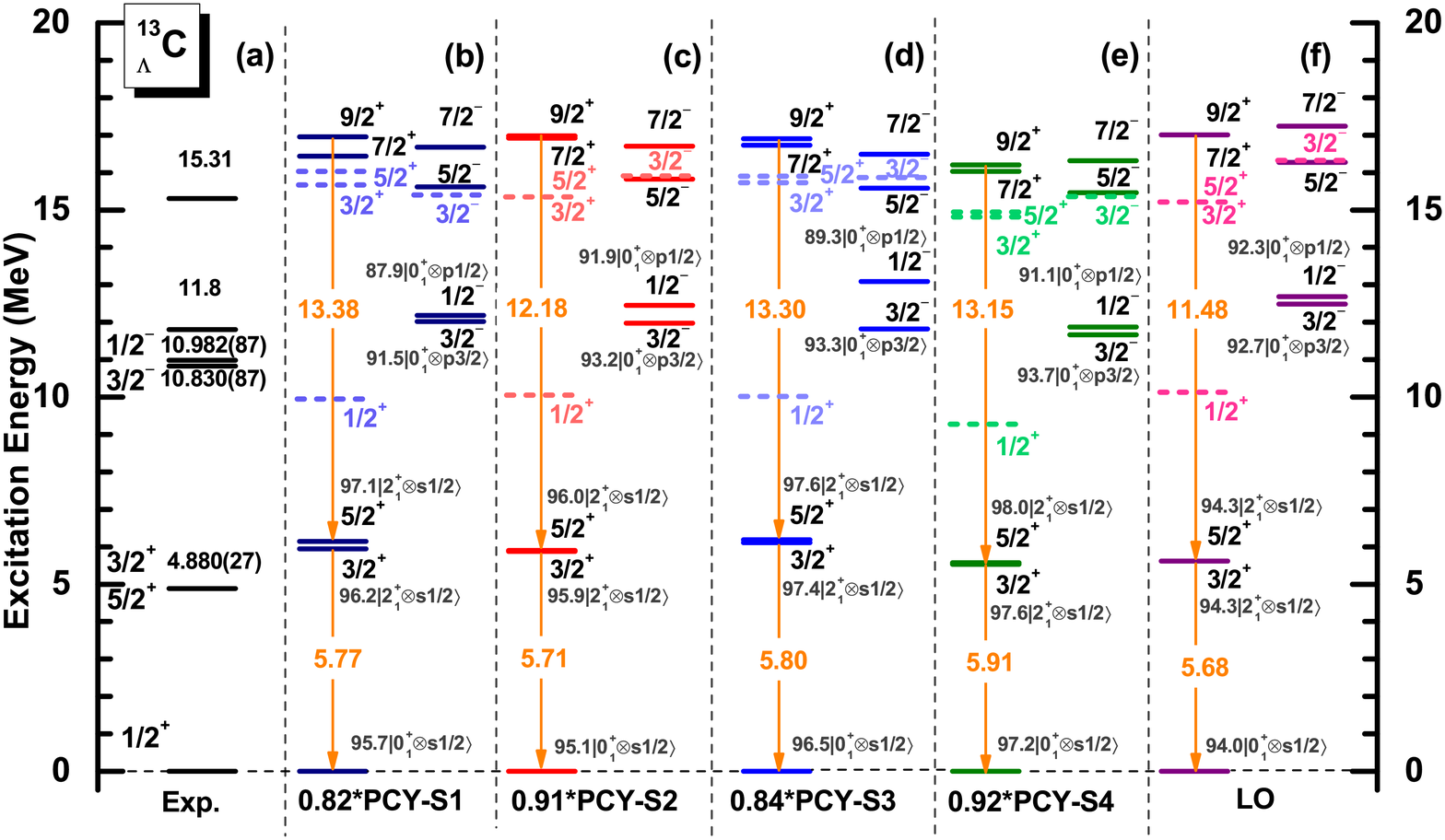}
 \caption{(Color online) Same as the Fig.~\ref{CSpectrum}, but with the scaled  $N\Lambda$
interaction, in which the scaling factor is determined for each parameter set to reproduce
the empirical $\Lambda$ binding energy of $^{13}_{~\Lambda}$C.}
  \label{C13L_2}
\end{figure*}

\section{Application to $^{13}_{~\Lambda}$C}
\label{Sec:Results}
Let us now apply the MPRM with the higher-order $N\Lambda$ interaction to  $^{13}_{~\Lambda}$C, for
which several low-lying states have been observed experimentally~\cite{Ajimura01,Kohri02}.
To this end, we first generate several low-lying states of the core nucleus $^{12}$C with a
quantum-number projected GCM calculation, where the mean-field configurations are obtained
from deformation constrained relativistic mean-field plus BCS calculation using the
point-coupling PC-F1 for the effective nucleon-nucleon interaction~\cite{Buvenich02}.
A zero-range pairing force supplemented with a smooth cutoff is adopted to treat the pairing
correlation among the nucleons. Axial symmetry and time-reversal invariance are imposed in the
mean-field calculations.
The Dirac spinor for each nucleon state is expanded on a harmonic oscillator basis with 10 shells.
More numerical details  have been presented in Refs.~\cite{Mei2014,Mei2015}.
The wave functions and the energies of the low-lying states of $^{12}$C are then used to calculate
the scalar and vector transition densities given by Eq. (\ref{TD}) as well as the matrix
elements in Eq. (\ref{Matrix_eq}). The radial wave function for the spherical harmonic oscillator
basis with 18 shells are used to expand the radial part of the
hypernuclear wave function, ${\mathscr R}_{k}(r)$.
We use the effective $N\Lambda$ interaction with the PCY-S1, PCY-S2, PCY-S3 and PCY-S4
parameter sets, which were determined by fitting to the experimental data of $\Lambda$
binding energies from light to heavy  mass hypernuclei~\cite{Tanimura2012}.
We list these parameters in Table~\ref{forces}.
Notice that PCY-S2 and PCY-S4 do not include the tensor and the derivative terms, respectively.
Notice also that PCY-S3 was obtained by excluding the spin-orbit splitting of the 1p state of
$\Lambda$ in $^{16}_{\Lambda}$O from the fitting, and the strength of the tensor coupling
is considerably smaller than that in PCY-S1.

\subsection{Low-energy spectra}

Figures ~\ref{CSpectrum}(b)-(e) show the calculated low-energy spectra of $^{13}_{~\Lambda}$C
with the higher order $N\Lambda$ interaction, in comparison with the experimental data as well
as with
the results of Ref.~\cite{Mei2015} obtained only with
the leading-order $N\Lambda$ interaction.
One can see that the calculated energy splitting between the $1/2^-$  and $3/2^-$ states,
as well as that between the $5/2^+$ and  $3/2^+$ states,
are clearly different among the four different parameter sets, although the main structures
of the low-lying states are the same.
That is, the splitting between the $1/2^-_1$ and $3/2^-_1$ with PCY-S1 and PCY-S4 forces are
smaller than that with PCY-S2 and PCY-S3 forces
and much close to the experiment data. The splitting between the $5/2^+$ and $3/2^+$ states by
the PCY-S1 are much larger than that by the other parameter sets.  In other words, the fine
structure of the hypernuclear low-lying states reflects the impact of the  $N\Lambda$
interaction beyond the leading order.
We have performed similar calculations for $^9_\Lambda$Be,
and have found that the
effects of the derivative and the tensor terms are similar to
those in $^{13}_\Lambda$C.
Notice that
even though the tensor term is absent in
the PCY-S2 force,
a good description is still acieved by
largely deviating from the expected relations of a naive quark
model, that is,
$\alpha^{N\Lambda} = \frac{2}{3}\alpha^{NN}$ etc. \cite{Toki94}.
We will further discuss the role of the higher order
terms in $N\Lambda$ interaction in the next subsections.
In particular, we will demonstrate that the tensor term
plays an important role if the expected relations of the naive
quark model are maintained.

In Fig.\ref{CSpectrum}, the $E2$ transition strengths between the low-lying states of
$^{13}_{~\Lambda}$C are also presented.
One can see that the $E2$ transition strengths do not much vary with the four  $N\Lambda$
effective interactions and are close to those with the LO interaction.

Given the fact that all the four parameter sets  of the effective $N\Lambda$ interaction
were adjusted to $\Lambda$ binding energy of hypernuclei at the mean-field
level~\cite{Tanimura2012}, the use of these forces in the present MPRM calculation
overestimates the $\Lambda$ binding energy of $^{13}_{~\Lambda}$C.
That is, the $\Lambda$ binding energy of $^{13}_{~\Lambda}$C defined as the energy difference
between the $0^+_1$ state of $^{12}$C and the $1/2^+_1$ state of $^{13}_{~\Lambda}$C
are calculated to be $15.72$, $13.63$, $15.42$ and $13.22$ MeV  using the PCY-S1, PCY-S2,
PCY-S3 and PCY-S4 sets of $N\Lambda$ interaction, respectively, while the empirical value
is $B^{\mathrm{exp}.}_{\Lambda}=11.38\pm0.05$ MeV\cite{Hashimoto06}.
If we want to reproduce the $\Lambda$ binding energy within this approach, we need to scale
all the coupling strengths in the parameters of the $N\Lambda$ interaction by $18\%$, $9\%$,
$16\%$ and $8\%$ for PCY-S1, PCY-S2, PCY-S3 and PCY-S4, respectively.

Figure~\ref{C13L_2} shows the calculated low-lying spectra of $^{13}_{~\Lambda}$C with those
scaled effective $N\Lambda$ interactions.
It is shown that the predicted low-lying excitation spectrum of $^{13}_{~\Lambda}$C is
slightly compressed
and the $E2$ transition strengths are somewhat increased.
On the other hand, the energy splitting between the $1/2^-_1$ and $3/2^-_1$ states remains
large by the PCY-S2 and PCY-S3 forces,
while it
is reduced from 303.7 keV (253.7 keV) to 161.5 keV(206.3 keV) after scaling the coupling
strengths for the PCY-S1 (PCY-S4) interaction.
Due to the slightly weaker $N\Lambda$ interaction, the  configuration mixing for the
$1/2^+_1$, $3/2^+_1$, $5/2^+_1$, $1/2^-_1$ and $3/2^-_1$ states becomes slightly reduced for
all the four parameter sets.

\subsection{Effects of the derivative coupling terms}

\begin{figure}[]
  \centering
 \includegraphics[width=8cm]{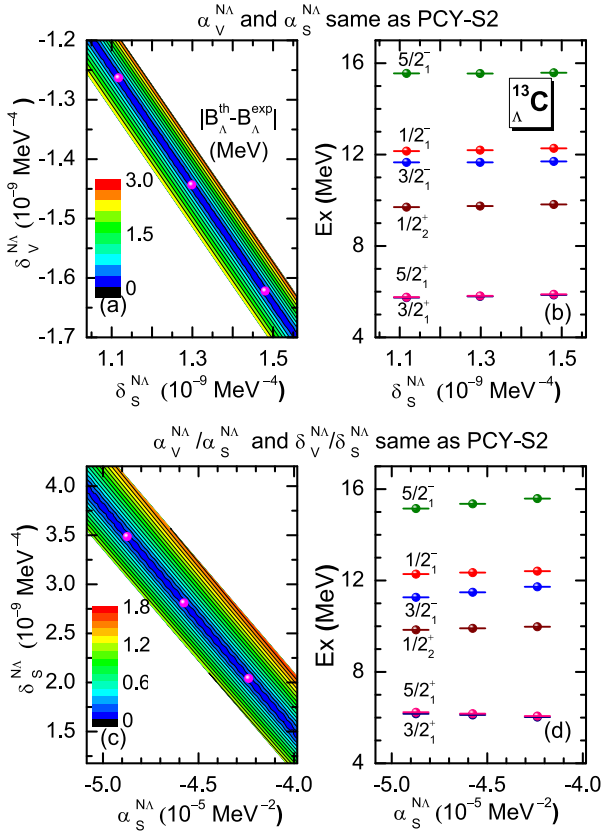}
 \caption{(Color online) (a) and (c): Contour plots for the absolute value of the difference
between the
theoretical and the experimental hyperon binding energies of $^{13}_{~\Lambda}$C hypernucleus
as a function of the coupling strength parameters ($\delta^{N\Lambda}_S$, $\delta^{N\Lambda}_V$)
and ($\delta^{N\Lambda}_S$, $\alpha^{N\Lambda}_S$), respectively.
In the former, $\alpha^{N\Lambda}_V$ and $\alpha^{N\Lambda}_S$ are fixed to the same values
as in PCY-S2,
while in the latter, the value of $\alpha^{N\Lambda}_V$ and $\delta^{N\Lambda}_V$
is determined for each
($\alpha^{N\Lambda}_S,\delta^{N\Lambda}_S$) so as to keep the ratios
$\alpha^{N\Lambda}_V/\alpha^{N\Lambda}_S$ and $\delta^{N\Lambda}_V/\delta^{N\Lambda}_S$
to be the same as those for PCY-S2.
(b) and (d): Low-lying states
in $^{13}_{~\Lambda}$C calculated with the  strength parameters
denoted by the dots in the panels (a) and (c), respectively. }
   \label{S2LODer}
\end{figure}

We now examine the effect of the derivative coupling terms on the $\Lambda$ binding energy.
To this end, we fix the
coupling strengths for the leading order terms ($\alpha^{N\Lambda}_V, \alpha^{N\Lambda}_S$) to be the same values as
those in the PCY-S2 force and study the $\Lambda$ binding energy as a function of the coupling
strengths ($\delta^{N\Lambda}_V$, $\delta^{N\Lambda}_S$) of the derivative terms.
Notice that the tensor coupling is absent in PCY-S2, so that we can isolate the effect of the
derivative terms.
The results are shown in Fig.~\ref{S2LODer}(a).
A clear linear correlation is observed between $\delta^{N\Lambda}_V$ and $\delta^{N\Lambda}_S$.
By selecting three sets of $(\delta^{N\Lambda}_V, \delta^{N\Lambda}_S)$ along the valley in
Fig.~\ref{S2LODer}(a),
we calculate the low-lying states of $^{13}_{~\Lambda}$C and show them  in Fig.~\ref{S2LODer}(b).
One can see that the low-lying states are similar to each other.
This implies that the coupling strengths $(\delta^{N\Lambda}_V, \delta^{N\Lambda}_S)$ may not
be uniquely determined by the energies
of hypernuclear low-lying states.

\begin{figure}[]
  \centering
 \includegraphics[width=8cm]{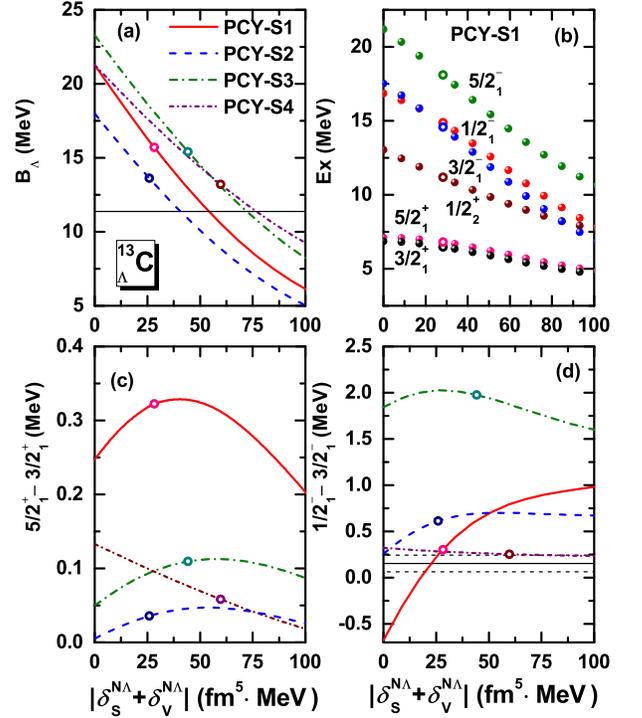}
 \caption{(Color online) (a): The $\Lambda$ binding energy in $^{13}_{~\Lambda}$C as a function
of $|\delta^{N\Lambda}_S+\delta^{N\Lambda}_V|$,
 while keeping the same values of $\alpha^{N\Lambda}_S$, $\alpha^{N\Lambda}_V$, $\alpha^{N\Lambda}_T$
and $\delta^{N\Lambda}_V/\delta^{N\Lambda}_S$ as the
 original values for the PCY-S1, PCY-S2,PCY-S3, and PCY-S4 parameter sets.
 $B_\Lambda$ with the original value of $\delta^{N\Lambda}_S$ and $\delta^{N\Lambda}_V$ is denoted by
the open circles for each parameter set.
 The experimental value is denoted by the thin solid line.
 (b): The energy levels of the low-lying states as a function of
$|\delta^{N\Lambda}_S+\delta^{N\Lambda}_V|$ for the PCY-S1 parameter set.
 (c) and (d): The energy splitting between the $5/2^+_1$ and $3/2^+_1$ states and that between
the $1/2^-_1$ and $3/2^-_1$ states, respectively,
as a function of $|\delta^{N\Lambda}_S+\delta^{N\Lambda}_V|$.
}
   \label{LODer}
\end{figure}
\begin{figure}[]
  \centering
 \includegraphics[width=8cm]{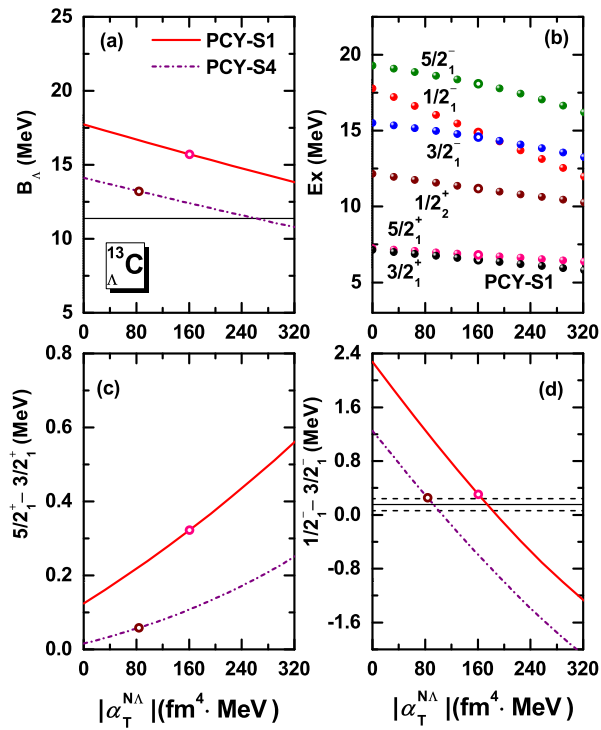}
 \caption{(Color online) Same as Fig.~\ref{LODer}, but as a function of the tensor
coupling strength
$|\alpha^{N\Lambda}_T| (=-\alpha^{N\Lambda}_T)$
 for the PCY-S1 and PCY-S4 forces.  }
  \label{S14Tensor}
\end{figure}

Since the vector coupling strengths $\alpha^{N\Lambda}_V$ and $\delta^{N\Lambda}_V$ are linearly
correlated with
the corresponding scalar coupling strengths $\alpha^{N\Lambda}_S$ and $\delta^{N\Lambda}_S$, respectively,
we next keep the ratios of $\alpha^{N\Lambda}_V/\alpha^{N\Lambda}_S$ and
$\delta^{N\Lambda}_V/\delta^{N\Lambda}_S$
to be the same as those in PCY-S2 force and calculate the
$\Lambda$ binding energy as well as the low-lying spectrum
as a function of $\alpha^{N\Lambda}_S$ and $\delta^{N\Lambda}_S$
as shown in Fig.~\ref{S2LODer}(c)
and (d), respectively.
It is shown that the parameters $\delta^{N\Lambda}_S$ and $\alpha^{N\Lambda}_S$ are also linearly
correlated when these are fitted to the $\Lambda$ binding energy in $^{13}_{~\Lambda}$C
(see Fig.~\ref{S2LODer}(c)).

Notice that the difference between the vector transition density $\rho^{nI n'I'}_{\lambda, V}(r)$
and the scalar transition density $\rho^{nI n'I'}_{\lambda, S}(r)$
in the low-lying states of $^{12}$C is small (see Fig.4 in Ref.~\cite{Mei2015}).
In the non-relativistic approximation, with the same scalar and vector densities,
the sum of LO coupling strengths $\alpha^{N\Lambda}_S+\alpha^{N\Lambda}_V$
and the sum of the derivative coupling strengths $\delta^{N\Lambda}_S+\delta^{N\Lambda}_V$ can
be regarded as the
depth of the central potential and the surface coupling strength, respectively.
Therefore, these are also linearly correlated, as has been found in Ref.~\cite{Hiyama14}.
Taking three sets of the parameters along the valley with
$B^{\mathrm{th}}_{\Lambda}=B^{\mathrm{exp}}_{\Lambda}$  in Fig.~\ref{S2LODer}(c),
we find that those three sets yield almost the same excitation energies (within around 0.13 MeV)
for the $3/2_1^+$, $5/2_1^+$, $1/2_2^+$ and $1/2^-_1$ states,
while the difference is much larger (around 0.45 MeV) for the $3/2_1^-$ and $5/2_1^-$ states.
A comparison between Figs~\ref{S2LODer}(b)and \ref{S2LODer}(d) suggests that the excitation
energies of the low-lying states
are more sensitive to $\alpha^{N\Lambda}_S$ and $\alpha^{N\Lambda}_V$ as compared to $\delta^{N\Lambda}_S$ and $\delta^{N\Lambda}_V$.

We next examine the influence of the derivative interaction terms for the other parameter
sets as well.
To this end, we vary $\delta^{N\Lambda}_S$ and $\delta^{N\Lambda}_V$ by keeping the values of
$\alpha^{N\Lambda}_S$, $\alpha^{N\Lambda}_V$, $\alpha^{N\Lambda}_T$ and the ratio
$\delta^{N\Lambda}_S/ \delta^{N\Lambda}_V$
to be the same as the original values for each parameter set.
Fig.~\ref{LODer}(a) shows the $\Lambda$ binding energy so obtained as a function
of $|\delta^{N\Lambda}_S+\delta^{N\Lambda}_V|=-(\delta^{N\Lambda}_S+\delta^{N\Lambda}_V)$.
The calculated $B_{\Lambda}$ with the original value of $\delta^{N\Lambda}_S$
and $\delta^{N\Lambda}_V$ is denoted by the open circle for each parameter set.
$B_{\Lambda}$  decreases with increasing $|\delta^{N\Lambda}_S+\delta^{N\Lambda}_V|$
and approaches to the experimental value denoted by the thin solid line.
The $\Lambda$ binding energy decreases from 21.28MeV
to 15.72 MeV  by adding the derivative coupling terms to the PCY-S1 interaction
(that is, by changing $|\delta^{N\Lambda}_S+\delta^{N\Lambda}_V|$ from 0 to the original value
denoted by the open cicle).
For PCY-S2, PCY-S3, and PCY-S4 interactions, the shift is from 18.01, 23.29, and 21.27 MeV
to 13.63, 15.42, and 13.22 MeV, respectively.

The excitation energies  of the low-lying states as a function of the derivative coupling
strength $|\delta^{N\Lambda}_S+\delta^{N\Lambda}_V|$
are shown in Figure~\ref{LODer}(b), where $\alpha^{N\Lambda}_S$, $\alpha^{N\Lambda}_V$ ,
$\alpha^{N\Lambda}_T$ and $\delta^{N\Lambda}_S/ \delta^{N\Lambda}_V$ are kept to
be the same as those for PCY-S1.
As one can see, the excitation energies decreases with the increase of
$|\delta^{N\Lambda}_S+\delta^{N\Lambda}_V|$.
Notice that the change of the $3/2^+$ and $5/2^+$ states are much smaller compared to the
change in the other states.
Similar behaviors are found also for the PCY-S2, PCY-S3 and PCY-S4 forces (not shown).
The energy splittings of ($3/2^+_1$, $5/2^+_1$) and ($1/2^-_1$, $3/2^-_1$) states as a
function of the strength of the derivative coupling terms are shown in Fig.~\ref{LODer}(c)
and (d), respectively.
It is found that the $5/2^+_1$ state is always slightly higher than  the $3/2^+_1$ state,
which is  by less than 0.15 MeV except for PCY-S1
in the range of $|\delta^{N\Lambda}_S+\delta^{N\Lambda}_V|$ shown in the figure.
In contrast, the available data indicate that the $5/2^+_1$ state is slightly lower
than the $3/2^+_1$ state.
This discrepancy may be due to the spin-spin $N\Lambda$ interaction~\cite{MCAS},
which is missing in the present calculations.

For the doublet of ($1/2^-,3/2^-$), the $1/2^-$ state is predicted to be higher than
the $3/2^-$ state for all the forces except for the PCY-S1,
with which the $1/2^-$ state is lower than the $3/2^-$ state
for $|\delta^{N\Lambda}_S+\delta^{N\Lambda}_V|<17.56$ MeV.
As will be discussed in the next subsection, this splitting, which reflects
the spin-orbit splitting of the $p_\Lambda$ hyperon~\cite{Mei2015},
is mainly governed by the tensor coupling term.

\subsection{Effects of the tensor coupling term}

Let us next examine the effects of the tensor coupling term on hypernuclear low-lying states.
For this purpose, we adopt the PCY-S1 and PCY-S4 sets of the $N\Lambda$  interaction and vary
the strength $\alpha^{N\Lambda}_T$ for the tensor coupling term.
Fig.~\ref{S14Tensor}(a) shows the $\Lambda$ binding energy of $^{13}_{~\Lambda}$C as a
function of $|\alpha^{N\Lambda}_T|=-\alpha^{N\Lambda}_T$.
The $\Lambda$ binding energy gradually decreases from 17.71 MeV (14.12 MeV) for
$\alpha^{N\Lambda}_T=0$ to 15.72 MeV (13.22 MeV)
for the original value of $\alpha^{N\Lambda}_T$ for the  PCY-S1 (PCY-S4) force, which
is indicated by the open circle in Fig.~\ref{S14Tensor}(a).

Figure~\ref{S14Tensor}(b) shows the excitation energies for the low-lying states of
$^{13}_{~\Lambda}$C as a function of the tensor coupling strength $|\alpha^{N\Lambda}_T|$ for
the PCY-S1. As already shown in the previous mean-field studies~\cite{Mares94,Weixia14,Toki94},
the tensor coupling term makes the $s_\Lambda$ hyperon less bound by increasing the energy of
the $s_{1/2}$ level.
Moreover, it decreases (increases) the energy of the hyperon $p_{3/2}$ ($p_{1/2}$)  state.
This is consistent with Fig.\ref{S14Tensor}(a) for the ground state ($1/2^+$) of  the
$^{13}_{~\Lambda}$C, the energy of which decreases by the tensor coupling. As a result, the
$\Lambda$ binding energy is reduced  by 0.9 MeV for the PCY-S1 and 1.99 MeV for the PCY-S4
after turning on the tensor coupling term. At the same time, the tensor coupling term decreases
(increases) the energy of the $3/2^-$ ($1/2^-$) state, which  mainly consists of
the $p_{3/2}$ ($p_{1/2}$) hyperon coupled to the ground state ($0^+$) of $^{12}$C.
Since the $1/2^-$ changes more significantly than the $3/2^-$ state, the higher
lying $1/2^-$  state approaches the $3/2^-$ state and even becomes lower than the $3/2^-$
state for large values of the tensor coupling strength,
indicating that the energy splitting of the $1/2^-$ and $3/2^-$ states is sensitive
to the tensor coupling strength.
For the PCY-S1 and PCY-S4 forces, the energy difference between $1/2^-_1$ and $3/2^-_1$ states
decreases from 2.28 MeV to 0.31 MeV, and from 1.25 MeV to 0.25 MeV,  respectively, while
turning on the tensor coupling term.
For the energy gap between the $3/2^+_1$ and $5/2^+_1$ states, it increases with increasing
the $|\alpha^{N\Lambda}_T|$, as shown in Fig.\ref{S14Tensor}(c). Again, the tensor coupling term does
not invert the energy ordering of the $3/2^+_1$ and $5/2^+_1$ states.

\section{Summary}
\label{Sec:Summary}

We have implemented the higher-order derivative and the tensor terms in the point coupling
$N\Lambda$ interaction in the microscopic particle-rotor model for hypernuclear low-lying states.
By taking $^{13}_{~\Lambda}$C as an example, we have adopted the four sets of effective
$N\Lambda$ interaction,
which were adjusted at the mean-field level to the $\Lambda$ binding energy.
We have shown that the four parameter sets yield a qualitatively similar low-lying spectrum
to one another,
even though these parameter sets were obtained using only the ground state energy.

We have discussed in detail the impact of each $N\Lambda$ interaction term on hypernuclear
low-lying states for $^{13}_{~\Lambda}$C.
We have shown that both the second-order derivative and the tensor coupling terms raise
the energy of hypernuclear states and thus reduce the $\Lambda$ binding energy.
With the increase of the tensor coupling strength, the
excitation energy of the $1/2^-$ state has been found to decrease
faster than the $3/2^-$ states.
As a result, the energy difference $E(1/2^-)-E(3/2^-)$ decreases to a small value and
even changes its sign for large values of the tensor coupling term.
We have also found that the energy ordering of the $3/2^+_1$ and $5/2^+_1$ states
cannot be reproduced by the present effective $N\Lambda$ interaction.
We note that the four-fermion coupling terms
$(\bar\psi_N \Gamma_i\psi_N)(\bar\psi_\Lambda \Gamma_i  \psi_\Lambda)$
with $\Gamma_i=\sigma_{\mu\nu}$ and $\gamma_\mu\gamma^5$, which provides the spin-spin
$N\Lambda$ interaction~\cite{Polinder06}, are not taken into account in the present study.
This interaction term may have an important influence on the energy ordering of the $3/2^+_1$
and $5/2^+_1$ states.
It will be interesting to study in near future the role of these terms in hypernuclear spectroscopy
with the present microscopic particle-rotor model.

Another interesting work is
to compare directly between the microscopic particle-rotor model and the generator coordinate
method for the whole $\Lambda$
hypernuclei~\cite{Mei15-2}
using the same point-coupling $N\Lambda$ interaction. A work is now in progress, and
we will report on it in a separate paper.

\section*{Acknowledgments}
This work was supported in part by the Tohoku University Focused Research Project \lq\lq
Understanding the origins for matters in universe\rq\rq, JSPS KAKENHI Grant Number 2640263,
the National Natural Science Foundation of China
under Grant Nos. 11575148,11475140,11305134, and the Fundamental Research Funds for the Central
University (XDJK2013C028).
\section*{Appendix A: The $N\Lambda$ effective interaction and the corresponding energy functional}

In this Appendix A, we show that the $N\Lambda$ interaction given by Eqs.~(\ref{Scalar}),
(\ref{Vector}) and (\ref{Tensor})
lead to the energy functional given by Eq.(\ref{EDF}).
The energy functional for $N\Lambda$ interaction is given by the expectation value of the
effective interaction $\hat{V}^{N\Lambda}$ at the Hartree level,
\begin{equation}
\label{interactionE}
E^{N\Lambda}_{\rm int}= \sum_{i=1}^{A_c}\int d\vec{r} d\vec{r}'   \psi^\dag_{\Lambda}(\vec{r})\psi^\dag_i(\vec{r}')
\hat{V}^{N\Lambda}(\vec{r},\vec{r}')
\psi_{\Lambda}(\vec{r})\psi_i(\vec{r}').
\end{equation}
Substituting the LO scalar effective interaction term,
\begin{equation}
\hat{V}^{N\Lambda}_{S}(\vec{r},\vec{r}')=\alpha_S^{N\Lambda} \gamma^0_{\Lambda}
\delta(\vec{r}-\vec{r}')\gamma^0_N
\end{equation}
to Eq.(\ref{interactionE}), one finds
\begin{align}
&E^{N\Lambda}_{\rm S}\nonumber \\
&= \sum_{i=1}^{A_c}\int d\vec{r} d\vec{r}'  \psi^\dag_{\Lambda}(\vec{r}) \psi^\dag_i(\vec{r}')
\alpha_S^{N\Lambda} \gamma^0_{\Lambda}\delta(\vec{r}-\vec{r}')\gamma^0_N
\psi_{\Lambda}(\vec{r})\psi_i(\vec{r}') \nonumber \\
&=  \alpha_S^{N\Lambda} \sum_{i=1}^{A_c}\int d\vec{r}
\psi^\dag_{\Lambda}(\vec{r})\gamma^0_{\Lambda}\psi_{\Lambda}(\vec{r})
\psi^\dag_i(\vec{r})\gamma^0_N\psi_i(\vec{r})\nonumber \\
&=   \int d\vec{r} \alpha_S^{N\Lambda} \rho^\Lambda_S(\vec{r})\rho_S(\vec{r}),
\end{align}
where $\rho_S$ and $\rho^{\Lambda}_S$ are the scalar densities defined as
\begin{equation}
\rho_S(\vec{r})=\sum_{i=1}^{A_c}\bar{\psi}_i(\vec{r})\psi_i(\vec{r}), ~~
\rho^{\Lambda}_S(\vec{r})=\bar{\psi}_{\Lambda}(\vec{r})\psi_{\Lambda}(\vec{r}).
\end{equation}
The effective interaction with the scalar derivative term,
\begin{align}
\hat{V}^{N\Lambda}_{\rm Der}(\vec{r},\vec{r}')=&\delta_S^{N\Lambda}\gamma^0_{\Lambda}
\Big[\overleftarrow{\nabla}^2\delta(\vec{r}-\vec{r}')+\delta(\vec{r}-\vec{r}')
\overrightarrow{\nabla}^2  \nonumber \\ ~&+2\overleftarrow{\nabla}\cdot
\delta(\vec{r}-\vec{r}')\overrightarrow{\nabla}\Big]\gamma^0_N,
\end{align}
leads to
\begin{align}
&E^{N\Lambda}_{\rm S}\nonumber \\
&= \sum_{i=1}^{A_c}\int d\vec{r} d\vec{r}'   \psi^\dag_{\Lambda}(\vec{r})\psi^\dag_i(\vec{r}')
\delta_S^{N\Lambda}\gamma^0_{\Lambda}
\Big[\overleftarrow{\nabla}^2\delta(\vec{r}-\vec{r}')\nonumber \\
&~~~
+\delta(\vec{r}-\vec{r}')\overrightarrow{\nabla}^2
+2\overleftarrow{\nabla}\cdot\delta(\vec{r}-\vec{r}')\overrightarrow{\nabla}\Big]\gamma^0_N
\psi_{\Lambda}(\vec{r})\psi_i(\vec{r}') \nonumber \\
&= \delta_S^{N\Lambda}\sum_{i=1}^{A_c}\int d\vec{r}
\Big\{[\nabla^2\psi^\dag_{\Lambda}(\vec{r})\gamma^0_{\Lambda}]\psi_{\Lambda}(\vec{r})
+[\psi^\dag_{\Lambda}(\vec{r})\gamma^0_{\Lambda}][\nabla^2 \psi_{\Lambda}(\vec{r})] \nonumber \\
&~~~
+2[\nabla \psi^\dag_{\Lambda}(\vec{r})\gamma^0_{\Lambda}] \cdot [\nabla \psi_{\Lambda}(\vec{r})] \Big\}
[\psi^\dag_i(\vec{r})\gamma^0_N \psi_i(\vec{r})] \nonumber \\
&= \int d\vec{r} \delta_S^{N\Lambda} \rho_S(\vec{r}) \nabla^2 \rho^\Lambda_S(\vec{r}).
\end{align}
A similar derivation holds also for the vector part of the $N\Lambda$ interaction.

On the other hand, the tensor effective interaction,
\begin{equation}
\hat{V}^{N\Lambda}_{T}(\vec{r},\vec{r}')=i\alpha_T^{N\Lambda}
\Big[\overleftarrow{\nabla}\cdot\vec{\gamma} \delta(\vec{r}-\vec{r}')
+\delta(\vec{r}-\vec{r}')\overrightarrow{\nabla}\cdot \vec{\gamma}  \Big]
\end{equation}
leads to
\begin{align}
&E^{N\Lambda}_{\rm T}\nonumber \\
&= \sum_{i=1}^{A_c}\int d\vec{r} d\vec{r}'   \psi^\dag_{\Lambda}(\vec{r}) \psi^\dag_i(\vec{r}')
i\alpha_T^{N\Lambda}
\Big[\overleftarrow{\nabla}\cdot\vec{\gamma} \delta(\vec{r}-\vec{r}')\nonumber \\
&~~~
+\delta(\vec{r}-\vec{r}')\overrightarrow{\nabla}\cdot \vec{\gamma}  \Big]
\psi_{\Lambda}(\vec{r}) \psi_i(\vec{r}')\nonumber \\
&= \alpha_T^{N\Lambda} \sum_{i=1}^{A_c} \int d\vec{r}
\Big\{[\nabla\psi^\dag_{\Lambda}(\vec{r})\gamma^0_{\Lambda}]
\cdot i\vec{\alpha} \psi_{\Lambda}(\vec{r})\nonumber \\
&~~~
+[\psi^\dag_{\Lambda}(\vec{r})\gamma^0_{\Lambda}][\nabla \cdot i\vec{\alpha}
\psi_{\Lambda}(\vec{r})]\Big\}
 [\psi^\dag_i(\vec{r})\psi_i(\vec{r})]\nonumber \\
&= \int d\vec{r} \alpha_T^{N\Lambda} \rho_V(\vec{r}) [\nabla \cdot
(\bar{\psi}_{\Lambda}(\vec{r})i\vec{\alpha} \psi_{\Lambda}(\vec{r}))]  \nonumber \\
&= \int d\vec{r} \alpha_T^{N\Lambda} \rho_V(\vec{r}) \rho^{\Lambda}_T(\vec{r}),
\end{align}
where $\rho_V$ and $ \rho^\Lambda_T$ are the vector and the tensor densities defined as
\bsub\begin{align}
\rho_V(\vec{r})=&\sum_{i=1}^{A_c}\psi^\dag_i(\vec{r})\psi_i(\vec{r}), \\
\rho^\Lambda_T(\vec{r})=&\nabla\cdot(\bar{\psi}_\Lambda(\vec{r})  i\vec{\alpha} \psi_{\Lambda}(\vec{r})).
\end{align}\esub
Putting all these together, we finally obtain Eq.(\ref{EDF}).

\section*{Appendix B: A derivation of Eq.(\ref{Vmatrix}) for the matrix
elements of the vector derivative coupling term}

With the $N\Lambda$ vector derivative effective interaction
$\displaystyle
\hat{V}_{D} =\delta_V^{N\Lambda}
\Big[\overleftarrow{\nabla}^2 \delta(\vec{r}-\vec{r}_i)
+ \delta(\vec{r}-\vec{r}_i)\overrightarrow{\nabla}^2+
2 \overleftarrow{\nabla}\cdot
\delta(\vec{r}-\vec{r}_i)
 \overrightarrow{\nabla}\Big]$, and the definition of
\begin{equation}
{\mathscr F}^{JM}_{j l I}(\hat{\vec{r}}, \{\vec{r}_i\})
=\sum_{m_I m}C^{JM}_{I m_I j m } {\mathscr Y}_{jlm}(\hat{\vec{r}})\Phi_{I m_I}(\{\vec{r}_i\}),
\end{equation}
where ${\mathscr Y}_{jlm}(\hat{\br})$ is the spinor spherical harmonics,
\begin{equation}
{\mathscr Y}_{j\ell m}(\hat{\vec{r}})=\sum_{m_l m_s }C^{j m}_{l m_l \frac{1}{2} m_s}
Y_{l m_l}(\vartheta,\varphi)\chi_{m_s},
\end{equation}
the coupling matrix element of the vector derivative term reads
\begin{align}
&\langle R^k_{\alpha l}(r){\mathscr F}^{JM}_{j l I}(\hat{\vec{r}}, \{\vec{r}_i\})
|\hat{V}_{D}|{\mathscr F}^{JM}_{j' l' I'}(\hat{\vec{r}}, \{\vec{r}_i\})R^{k'}_{\alpha' l'}(r)\rangle
\nonumber \\
=& \delta_S^{N\Lambda}\sum_{m_{I}m} \sum_{m'_{I}m'} C^{J M}_{I m_I j m }C^{JM}_{I' m'_I j' m'} \nonumber \\
~&\times\sum_{\lambda\mu}\int r^2 dr \int d \hat{\vec{r}} \langle  \Phi_{Im_I}|\sum_{i=1}^{A_c}
\frac{\delta(r-r_i)}{r r_i}Y_{\lambda\mu}(\hat{\vec{r}}_i)|\Phi_{I'm'_I}\rangle \nonumber \\
~&\times Y_{\lambda\mu}^*(\hat{\vec{r}})
\Delta [{\mathscr Y}_{j\ell m}^*(\hat{\vec{r}}){\mathscr Y}_{j'\ell' m'}(\hat{\vec{r}})
R^k_{\alpha l}(r)R^{k'}_{\alpha' l'}(r)].
\end{align}
Here, we notice
\begin{align}
&\langle  \Phi_{Im_I}|\sum^{A_c}_{i=1}\frac{\delta(r-r_i)}{r r_i}
Y_{\lambda\mu}(\hat{\vec{r}}_i)|\Phi_{I'm'_I}\rangle  \nonumber \\
=& (-1)^{I-m_I}
\left(\begin{matrix}
I      &\lambda   & I'    \\
-m_I   &\mu       & m'_I
\end{matrix}\right)
\langle  \Phi_{I}||\sum^{A_c}_{i=1}\frac{\delta(r-r_i)}{r r_i}Y_{\lambda}(\hat{\vec{r}}_i)|\Phi_{I'}
\rangle \nonumber \\
=& (-1)^{I-m_I}
\left(\begin{matrix}
I      &\lambda   & I'    \\
-m_I   &\mu       & m'_I
\end{matrix}\right) \rho^{II'}_{\lambda,V}(r).
\end{align}
With the relation of
\begin{align}
&{\mathscr Y}_{j\ell m}^*(\hat{\vec{r}}){\mathscr Y}_{j'\ell' m'}(\hat{\vec{r}}) \nonumber \\
=&\sum_{m_l m_s}\sum_{m'_l m'_s}
C^{jm}_{l m_l \frac{1}{2} m_s} C^{j'm'}_{l' m'_l \frac{1}{2} m'_s}
\delta_{m_s m'_s}(-1)^{m_l} \nonumber \\
~&\times
\sum_{L M} \frac{\hat{l}\hat{l}'}{\sqrt{4\pi} \hat{L}}
C^{L0}_{l 0 l' 0}C^{LM}_{l -m_l l' m'_l}Y_{L m_L} (\hat{\vec{r}}),
\end{align}
we have
\begin{align}
&\Delta [{\mathscr Y}_{j\ell m}^*(\hat{\vec{r}}){\mathscr Y}_{j'\ell' m'}(\hat{\vec{r}})]\nonumber \\
&=\sum_{m_l m_s}\sum_{m'_l m'_s}\sum_{L M} \frac{\hat{l}\hat{l}'}{\sqrt{4\pi} \hat{L}}
C^{jm}_{l m_l \frac{1}{2} m_s} C^{j'm'}_{l' m'_l \frac{1}{2} m'_s}
\delta_{m_s m'_s}(-1)^{m_l} \nonumber \\
~&\times
C^{L0}_{l 0 l' 0}C^{LM}_{l -m_l l' m'_l}
\left[\frac{1}{r^2}\frac{d}{dr}\left(r^2\frac{d}{dr}\right)-\frac{L(L+1)}{r^2}\right]
Y_{L m_L} (\hat{\vec{r}}).
\end{align}
According to the orthogonalization of the spherical harmonics,
\begin{equation}
\int Y_{\lambda\mu}^*(\hat{\vec{r}})Y_{L m_L} (\hat{\vec{r}}) d \hat{\vec{r}}
=\delta_{\lambda,L}\delta_{\mu,m_L},
\end{equation}
the matrix element is then given by
\begin{align}
&\langle R^k_{\alpha l}(r){\mathscr F}^{JM}_{j l I}(\hat{\vec{r}}, \{\vec{r}_i\}) |\hat{V}_{D}|
{\mathscr F}^{JM}_{j' l' I'}(\hat{\vec{r}}, \{\vec{r}_i\})R^{k'}_{\alpha' l'}(r)\rangle \nonumber \\
=&\delta_V^{N\Lambda} (-1)^{j'+I+J}\sum_{\lambda}
\left\{\begin{matrix}
 J  & I   &l     \\
 \lambda   & j'        &I'
\end{matrix}\right\}
\langle j\ell|| Y_{\lambda} || j'\ell' \rangle \nonumber \\
&~ \times\int r^2 dr  \rho^{II'}_{\lambda,V}(r)  \left[\frac{1}{r^2}\frac{d}{dr}\left(r^2\frac{d}{dr}
\right)-\frac{\lambda(\lambda+1)}{r^2}\right]\nonumber \\
&~ \times
[R^k_{\alpha l}(r)R^{k'}_{\alpha' l'}(r)].
\end{align}

\section*{Appendix C: A derivation of Eq.(\ref{Tmatrix}) for the matrix
elements of the tensor coupling term}

The matrix elements of the tensor coupling term is given by
\begin{align}
T_{\alpha\alpha'}^{kk'}&\equiv\langle R^k_{\alpha l}(r){\mathscr F}^{JM}_{j l I}(\hat{\vec{r}},
\{\vec{r}_i\}) |
\alpha_T^{N\Lambda}\sum_{i=1}^{A_c}
\Big[\overleftarrow{\nabla} \delta(\vec{r}-\vec{r}_i)\nonumber \\
&~
+\delta(\vec{r}-\vec{r}_i)\overrightarrow{\nabla}\Big]\cdot \vec{\sigma}
|{\mathscr F}^{JM}_{j' \tilde{l}' I'}(\hat{\vec{r}}, \{\vec{r}_i\}) R^{k'}_{\alpha' \tilde{l}'}(r)\rangle
\nonumber \\
&=\alpha^{N\Lambda}_T \sum_{m'_{I}m'} \sum_{m_{I}m}
C^{J M}_{I' m'_I j' m'}C^{JM}_{I m_I j m}  \nonumber \\
&~\times\sum_{\lambda\mu}\int r^2 dr d\hat{\vec{r}}
\langle  \Phi_{Im_I}|\sum_i^{A_c}\frac{\delta(r-r_i)}{r r_i}Y_{\lambda\mu}(\hat{\vec{r}}_i)|\Phi_{I'm'_I}
\rangle  \nonumber \\
&~\times Y_{\lambda\mu}^*(\hat{\vec{r}}) \nabla \cdot [R^{k*}_{\alpha l}(r)
{\mathscr Y}^*_{jlm}(\hat{\vec{r}})\vec{\sigma}
R^{k'}_{\alpha' \tilde{l}'}(r){\mathscr Y}_{j'\tilde{l}'m'}(\hat{\vec{r}})].
\end{align}
Notice
\begin{align}
&\nabla \cdot [R^{k*}_{\alpha l}(r){\mathscr Y}^*_{jlm}(\hat{\vec{r}})\vec{\sigma}
R^{k'}_{\alpha' \tilde{l}'}(r){\mathscr Y}_{j'\tilde{l}'m'}(\hat{\vec{r}})] \nonumber \\
&=\Big[-\frac{d R^k_{\alpha l}(r)}{dr}-\frac{\kappa+1}{r}R^k_{\alpha l}(r)\Big]
[R^{k'}_{\alpha' \tilde{l}'}(r){\mathscr Y}^*_{j\tilde{l}m}(\hat{\vec{r}}){\mathscr Y}_{j'\tilde{l}'m'}(\hat{\vec{r}})]
\nonumber \\
&-\Big[\frac{d R^{k'}_{\alpha' \tilde{l}'}(r)}{dr}-\frac{\kappa'-1}{r}R^{k'}_{\alpha' \tilde{l}'}(r)\Big]
[R^{k*}_{\alpha l}(r){\mathscr Y}^*_{jlm}(\hat{\vec{r}}){\mathscr Y}_{j'l'm'}(\hat{\vec{r}})].
\end{align}
With the Wigner-Eckart theroem, one obtains
\begin{eqnarray}
&&\int d\hat{\vec{r}} {\mathscr Y}^*_{jlm}(\hat{\vec{r}}) Y_{\lambda\mu}^*(\hat{\vec{r}})
{\mathscr Y}_{j'l'm'}(\hat{\vec{r}}) \nonumber \\
&&= (-1)^{\mu+j-m}
\left(\begin{matrix}
j      &\lambda   & j'    \\
-m     &-\mu      & m'
\end{matrix}\right)
\langle jl || Y_{\lambda} ||j'l' \rangle.
\end{eqnarray}
From the relation
\begin{align}
&\sum_{m'_{I}m'} \sum_{m_{I}m} \sum_{\mu}C^{J M}_{I' m'_I j' m'}C^{JM}_{I m_I j m}
 (-1)^{I-m_I}
\left(\begin{matrix}
I      &\lambda   & I'    \\
-m_I   &\mu       & m'_I
\end{matrix}\right)\nonumber \\
&\times(-1)^{\mu+j-m}
\left(\begin{matrix}
j      &\lambda   & j'    \\
-m     &-\mu      & m'
\end{matrix}\right) =(-1)^{I'+J+j}\left\{\begin{matrix}
J        & I     & j   \\
\lambda  & j'    & I'
\end{matrix}\right\},
\end{align}
one finally obtains
\begin{align}
T_{\alpha\alpha'}^{kk'} &= -\alpha^{N\Lambda}_T(-1)^{j+I'+J}\sum_{\lambda}
\left\{\begin{matrix}
J        & I     & j   \\
\lambda  & j'    & I'
\end{matrix}\right\}\int r^2 dr \rho^{I'I}_{\lambda,V}(r) \nonumber \\
&~\times\Big\{ \Big[\frac{d R^k_{\alpha l}(r)}{dr}+\frac{\kappa+1}{r}R^k_{\alpha l}(r)\Big]
R^{k'}_{\alpha' \tilde{l}'}(r)
\langle j\tilde{l} || Y_{\lambda} ||j'\tilde{l}'\rangle \nonumber \\
&~+\Big[\frac{d R^{k'}_{\alpha' \tilde{l}'}(r)}{dr}-\frac{\kappa'-1}{r}R^{k'}_{\alpha' \tilde{l}'}(r)\Big]
R^k_{\alpha l}(r)
\langle jl || Y_{\lambda} ||j'l' \rangle
\Big\}.
\end{align}



\begin{thebibliography}{99}{}
\bibitem{Hashimoto06} O. Hashimoto and H. Tamura, Prog. Part. Nucl. Phys.
\textbf{57}, 564 (2006).

 \bibitem{Tamura09} H. Tamura, Int. J. Mod. Phys. A {\bf 24}, 2101 (2009).

\bibitem{Glendenning00} N. Glendenning, {\it Compact Stars} (Springer-Verlag, New York, 2000).

\bibitem{abinitio}
R. Wirth, D. Gazda, P. Navratil, A. Calic, J. Langhammer, and
R. Roth, Phys. Rev. Lett. \textbf{113}, 192502 (2014).

\bibitem{Motoba83}T. Motoba, H. Band\={o}, and K. Ikeda, Prog. Theor. Phys. \textbf{70}, 189 (1983).

\bibitem{Hiyama99} E. Hiyama, M. Kamimura, K. Miyazaki, and T. Motoba,
Phys. Rev. C \textbf{59}, 2351 (1999).

\bibitem{Bando90}  H. Bando, T. Motoba and J. \v{Z}ofka,
Int. J. Mod. Phys. \textbf{A 5}, 4021 (1990).

\bibitem{Hiyama03} E. Hiyama, Y. Kino, and M. Kamimura,
Prog. Part. Nucl. Phys. \textbf{51}, 223 (2003).

\bibitem{Cravo02} E. Cravo, A. C. Fonseca, Y. Koike,
Phys. Rev. C \textbf{66}, 014001 (2002).

\bibitem{Suslov04} V. M. Suslov, I. Filikhin, and B. Vlahovic,
J. Phys. G: Nucl. Part. Phys. \textbf{30}, 513 (2004).

\bibitem{Shoeb09}M. Shoeb and Sonika, Phys. Rev. C \textbf{79}, 054321 (2009).

\bibitem{Dalitz78} R. H. Dalitz and A. Gal, Ann. Phys. (N.Y.)
\textbf{116}, 167 (1978).

\bibitem{Gal71}A. Gal, J.M. Soper, and R.H. Dalitz,
Ann. Phys. (N.Y.) {\bf 63}, 53 (1971).

\bibitem{Millener}D. J. Millener, Nucl. Phys. {\bf A804}, 84 (2008);
{\bf A914}, 109 (2013).

\bibitem{Isaka11} M. Isaka, M. Kimura, A. Dot\'e and A. Ohnishi,
Phys. Rev. C \textbf{83}, 044323 (2011).

\bibitem{Isaka11-2}
M. Isaka, M. Kimura, A. Dot\'e and A. Ohnishi,
Phys. Rev. C \textbf{83}, 054304 (2011).

\bibitem{Isaka12}
M. Isaka, H. Homma, M. Kimura, A. Dot\'e and A. Ohnishi,
Phys. Rev. C \textbf{85}, 034303 (2012).

\bibitem{Isaka13}
M. Isaka, M. Kimura, A. Dot\'e and A. Ohnishi,
Phys. Rev. C \textbf{87}, 021304(R) (2013).

\bibitem{Zhou07} X. R. Zhou ,
H.-J. Schulze, H. Sagawa, C. X. Wu, and E.-G. Zhao,
Phys. Rev. C \textbf{76}, 034312 (2007).

\bibitem{Win08} M. T. Win and K. Hagino, Phys. Rev. C \textbf{78}, 054311
(2008).

\bibitem{Schulze10} H.-J. Schulze, M. T. Win, K. Hagino, and H. S. Sagawa,
Prog. Theo. Phys. \textbf{123}, 569 (2010).

\bibitem{Win11} Myaing Thi Win, K. Hagino, and T. Koike,
Phys. Rev. C \textbf{83}, 014301 (2011).

\bibitem{Lu11} B.-N. Lu, E.-G. Zhao, and S.-G. Zhou,
Phys. Rev. C \textbf{84}, 014328 (2011).

\bibitem{Weixia14} W. X. Xue, J. M. Yao, K. Hagino, Z. P. Li, H. Mei, and Y. Tanimura,
Phys. Rev. C \textbf{91}, 024327 (2015).

\bibitem{Li13} A. Li, E. Hiyama, X.-R. Zhou, and H. Sagawa,
Phys. Rev. C \textbf{87}, 014333 (2013).

\bibitem{Lu14} B.-N. Lu, E. Hiyama, H. Sagawa, and S.-G. Zhou,
Phys. Rev. C \textbf{89}, 044307 (2014).

\bibitem{HY14} K. Hagino and J.M. Yao,
in {\it Relativistic Density Functional for Nuclear Structure},
Int. Rev. Nucl. Phys. \textbf{10}, 263-303 (2016),
edited by J. Meng (World Scientific, Singapore, 2016).

\bibitem{Mei15-2} H. Mei, K. Hagino,  and J. M. Yao,
Phys. Rev. C \textbf{93},  011301(R) (2016).

\bibitem{Mei2014} H. Mei, K. Hagino, J. M. Yao, and T. Motoba,
Phys. Rev. C \textbf{90},  064302 (2014).

\bibitem{Mei2015} H. Mei, K. Hagino, J. M. Yao, and T. Motoba,
Phys. Rev. C \textbf{91}, 064305  (2015).

\bibitem{Tanimura2012} Y. Tanimura and K. Hagino,
Phys. Rev. C \textbf{85}, 014306 (2012).

\bibitem{Hiyama14} E. Hiyama, Y. Funaki, N. Kaiser, and W. Weise,
Prog. Theor. Exp. Phys., \textbf{2014}, 013D01(2014).

\bibitem{Noble80} J. V.Noble, Phys. Lett. B \textbf{89}, 325,(1980).

\bibitem{Buvenich02} T. Burvenich, D. G. Madland, J. A. Maruhn, and P.-G. Reinhard,
Phys. Rev. C \textbf{65}, 044308 (2002).

\bibitem{Ring80}
P. Ring and P. Schuck, {\it The Nuclear Many-Body Problem}
(Springer-Verlag, Berlin, 1980).

\bibitem{Polinder06} H. Polinder, J. Haidenbauer, Ulf-G. Meissner, Nucl. Phys. {\bf A779}, 244 (2006).

\bibitem{Yao15} J. M. Yao, M. Bender, and P.-H. Heenen,
Phys. Rev. C {\bf 91}, 024301 (2015).

\bibitem{Ajimura01} S. Ajimura {\it et al}.,
Phys, Rev. Lett. {\bf 86},4255 (2001).

\bibitem{Kohri02}H. Kohri {\it et al}.,
Phys.Rev.C {\bf 65},034607 (2002).

\bibitem{Toki94} Y. Sugahara, and H. Toki, Prog. Theo. Phys. \textbf{92}, 803(1994).

\bibitem{MCAS}L. Canton, K. Amos, S. Karataglidis, and J. P. Svenne, Int. J.
Mod. Phys. E {\bf 19}, 1435 (2010).

\bibitem{Mares94} J. Mare$\check{\mathrm{s}}$ and B. K. Jenings,
Phys. Rev. C \textbf{49}, 2472-2478 (1994).

\end{thebibliography}
\end{document}